\newcommand{\be}{\begin{equation}}
\newcommand{\ee}{\end{equation}}
\newcommand{\bea}{\begin{eqnarray}}
\newcommand{\eea}{\end{eqnarray}}

\documentclass[epj]{svjour}
\usepackage{graphics}
\begin{document}
\title{Strength of singularities in varying constants theories}
\author{Konrad Marosek\inst{1} \and Adam Balcerzak\inst{2,3}
}                     
%
%
\institute{Chair of Physics, Maritime University, Wa{\l }y Chrobrego 1-2, 70-500 Szczecin, Poland \and Institute of Physics, University of Szczecin,  Wielkopolska 15, 70-451 Szczecin,  Poland \and Copernicus Center for Interdisciplinary Studies,   S{\l }awkowska 17, 31-016 Krak\'ow, Poland}
\date{Received: date / Revised version: date}
%
\abstract{
In this paper we consider a specific type of the bimetric theory of gravitation with the two different metrics introduced in the cosmological frame. Both metrics respect all the symmetries of the standard FLRW  solution and contain conformally related spatial parts. One of the metric is assumed to describe the causal structure for the matter. Another metric defines the causal structure for the gravitational interactions. A crucial point is that the spatial part of the metric describing gravity is given by the spatial part of the matter metric confromally rescaled by a time-dependent factor $\alpha$ which, as it turns out, can be linked to the effective gravitational constant and the effective speed of light. In the context of such a bimetric framework we examine the strength of some singular cosmological scenarios in the sense of the criteria introduced by Tipler and Kr\'olak. In particular, we show that for the nonsingular scale factor associated with the matter metric, both the vanishing or blowing up of the factor $\alpha$ for some particular moment of the cosmic expansion may lead to a strong singularity with infinite value of the energy density and infinite value of the pressure.
} 
\maketitle
\section{Introduction}
\label{sec:intro}

The number of the observed phenomena explained within the framework of the standard cosmological model is growing constantly which may indicate that it is entering the phase of achieving its final and complete from. The standard model properly describes the history of the universe starting form the era of inflation up to the present moment. However, the era containing the  initial singularity has  not yet been properly incorporated into the framework of the standard cosmological model. So far many attempts have been made in order to include the initial singularity into the cosmological framework. Among them one should mention the ekpyrotic \cite{Khoury,Khoury2}, the cyclic \cite{Steinhardt,Steinhardt2} and the pre-big-bang scenario \cite{Gasperini} - each one based on the tree level low-energy-effective action of the string theory - introducing pre-big-bang eras and this way somehow circumventing the problem of initial singularity. On the other hand, the discovery of the accelerated expansion of the late time Universe, and the fact that the observational data is insufficient to discriminate between different models of the dark energy encouraged many authors to speculate about the future evolution of the Universe. In particular, models based on phantom matter yet not ruled out by observations lead to scenarios terminating with a Big-Rip singularity \cite{Caldwell}. Relaxation of the barotropic constraints on the dark energy equation of state gave raise to many different scenarios containing future singularities like a sudden future singularity (SFS or type II) \cite{Barrow}, generalized sudden future singularities (GSFS) \cite{GSFS}, finite scale factor singularities (FSF or type III) \cite{Odintsov1,FSF}, big-separation singularities (BS or type IV) \cite{Odintsov2}, and w-singularities \cite{w}, little-rip and pseudo-rip singularities \cite{Frampton,Frampton1}.
Since all of the abovementioned singularities  are curvature singularities and prove their geodesic completeness (with exception for the Big-Rip) the other criterions such as those proposed by Tipler and Kr\'olak relying on the notion of the strength of singularity \cite{TiplerDef,KrolakDef} have to be used in order to differentiate between them. As it was founded in \cite{Marosek1,Marosek2}, some of those singularities can be weaken by assuming the variation of fundamental constants.

The aim of this paper is to use a variant of the bimetric gravity theory proposed in \cite{Clayton} to investigate the conditions in which some of the curvature singularities may occur. We will use two different metrics specified in the cosmological frame with conformally related spatial parts. One of the metric introduced defines the causal structure for the matter, and the other specifies the causal structure for the gravitational interactions. We will show that even if the behaviour of the matter metric is perfectly regular, the appearance of the strong singularity in the gravitational metric in the sense of Tipler and Kr\'olak may imprint itself in the singular behaviour of the matter. Additionally, in the presented model, the particular form of the conformal relation between the spatial parts of both metrics determine the value of the effective gravitational constant and the effective speed of light.
The study on the cosmological scenarios containing Big Rip and Little Rip singularities in the context of bimetric theories
was performed in \cite{Odintsov3}. However, the question of the strength of the singularities and their appearance in theories with varying fundamental
constants was not raised.

The paper is organized as follows. In Section \ref{sec2} we discus the peculiarities of the bimetric framework used in our paper. In Section \ref{sec3} we show that a singular behaviour of the matter fields may be associated with the occurrence of the strong singularity in the gravitational metric which defines the causal structure for gravitational interactions.  In Section \ref{sec4} we give our conclusions.

\section{Dynamical fundamental constants in bimetric approach}
\label{sec2}

As it was pointed out in \cite{Clayton}, the bimetric theories are capable to include  cosmological scenarios with varying speed of light. In such theories the light follows the causal structure defined by the matter metric, so the speed of light may be varying from the point of view of the spacetime with the causal structure defined by the gravitational metric. In this paper, we will follow a similar scheme and introduce two different metrics - the gravitational metric ${\hat{g}}_{\mu \nu}$ that specifies the causal structure for the gravitational field, and the matter metric $g_{ \mu \nu}$ that defines the causal structure for the mater fields. It should be stressed that the formulation of our model is restricted only to the diagonal spacetimes.

The relation between these metrics can be expressed by:
\be
\label{MetricForm1}
{\hat{g}}_{\mu \nu}= g_{\mu \nu} {\left[ \alpha - \left( \alpha - 1 \right) \left( {\delta}_{ 0 \mu} {\delta}_{ 0 \nu} \right) \right]}^2 ~,
\ee
where $ \alpha = \alpha \left( t \right) $ is a dimensionless time dependent function with the individual elements of the metric tensor given by:
\be \label{g00}
{\hat{g}}_{0 0}=g_{0 0}~,
\ee
\be \label{g11}
{\hat{g}}_{1 1}= {\alpha}^2 g_{1 1}~,
\ee
\be \label{g22}
{\hat{g}}_{2 2}= {\alpha}^2 g_{2 2}~,
\ee
\be \label{g33}
{\hat{g}}_{3 3}= {\alpha}^2 g_{3 3}~.
\ee
The time components in both metrics are identical while spatial elements of the matter metric are scaled by function $\alpha^2$ in comparison with the elements of the gravitational metric. We will see that in the considered model, the dynamical character of the relation between the two metrics will enable variation of the speed of light and the gravitational constant.
We assume that the expansion of the Universe, as seen from the perspective of the matter frame, can be different than that seen from the perspective of the gravitational frame. Consequently, the strength of any given singular scenario may depend upon the chosen perspective. The relation between the two metric also implies that the expansion may influence the propagation of the gravitational waves in a different way than it does influence the propagation of light (the delay between gravitational waves and the light was measured in binary neutron stars \cite{Ligo,Ligo2}).
The relation (\ref{MetricForm1}) arise naturally in the theory of disformal gravity \cite{Bekenstein} which is a specific type of bimetric gravity theory based on the most general mapping between the two metrics involving one scalar field and preserving diffeomorphisms of spacetime. The usual form of the mapping is given by:
\be
\label{DisformalMetric}
{\tilde{g}}_{\mu \nu}= C \left( \phi , X \right) g_{\mu \nu} + D \left( \phi, X \right) {\partial}_{\mu} \phi {\partial}_{\nu} \phi ~,
\ee
where $X = {\partial}_{\mu} \phi {\partial}^{\mu} \phi$ and $\phi$ is a kinetic term of the scalar field $\phi$, and  $C \left( \phi , X \right)$ and $D \left( \phi, X \right)$ are some general functions of $\phi$ and $X$. In order to see that (\ref{MetricForm1}) can be obtained from (\ref{DisformalMetric}) we choose $C \left( \phi , X \right)\equiv\phi^2$ and $D \left( \phi, X \right) = f \left( \phi \right) + g \left( \phi \right) {\partial}_{\mu} \phi {\partial}^{\mu} \phi$ where $f(\phi)$ and $g(\phi)$ are some not yet specified functions of the scalar field $\phi$. The choice of the homogeneous and isotropic cosmological frame (this can only be realized by assuming Friedmann-Lamaitre-Robertson-Walker line element) implies the spatial constancy of the scalar field which in the expanding setup gives that $\phi=\phi(t)$. The two assumptions above lead us to the following relation between the diagonal (the only non-vanishing in the considered model) components of the two metrics:
\bea
\label{relg00}
{\tilde{g}}_{0 0} &=& {\phi}^2 g_{0 0} + f \left( \phi \right) {\dot{\phi}}^2 + g \left( \phi \right) {\dot{\phi}}^4 g^{0 0}~, \\
\label{relgkk}
{\tilde{g}}_{k k} &=& { \phi}^2 g_{k k}~.
\eea
The formula (\ref{relgkk}) already gives the relation between spatial components of the two metrics of the type given by Eqs. (\ref{g11})-(\ref{g33}). Assumption of $g_{00}=g^{00}=-1$ (this can always be done by choosing the time coordinate $x^0$ to be a proper time of the comoving observers) leads to the relation between the time components of the two metrics of the type given by (\ref{g00}), provided that
\be \label{govphi}
-{\phi}^2 + f \left( \phi \right) { \dot{\phi} }^2 - g \left( \phi \right) { \dot{\phi} }^4 = -1~.
\ee

The total action can be written as a sum of the gravitational action and matter field action
\be
\label{Action}
S= S_g + S_{matter}~,
\ee
where the action for the gravitation is
\be
\label{gravaction}
S_g= -\frac{1}{16 \pi G_0} \int d^4 x R \left[ \hat{g} \right] \sqrt{-\hat{g}} ~.
\ee
The gravitational action is identical with  the standard Einstein-Hilbert action excluding cosmological term $ \Lambda = 0 $. We assume that this action is calculated on the basis of the gravitational metric $ {\hat{g}}_{ \mu \nu } $. Here $ G_0 $ is a constant with the same unit as the Newton constant $ G $, but different in its value. The action for the matter field is given by
\be
\label{fieldaction}
S_{matter}=-\frac{1}{2 {c_0}} \int d^4 x L_{matter} \sqrt{-g}~,
\ee
Bearing in mind that $\sqrt{-{\hat{g}}}= {\alpha}^{3} \sqrt{-g}$, the variation of the action (\ref{Action}) with respect to  $g^{ \mu \nu }$ gives
\bea \nonumber
-\frac{{c_0}^3}{16 \pi G_0} \int d^4 x \left[ \frac{{\alpha}^{3}}{{\left[ \alpha - \left( \alpha - 1 \right) \left( {\delta}_{ 0 \mu} {\delta}_{ 0 \nu} \right) \right]}^2}\right. \times\\ \left.
 \times \left(R_{\mu \nu} \left[\hat{g}\right] - \frac{1}{2} {\hat{g}}_{\mu \nu} R \left[\hat{g}\right] \right) - \frac{8 \pi G_0}{c_0^4} T_{\mu \nu} \right] \sqrt{-g} \delta g^{\mu \nu} = 0 ~,
\eea
which yields the field equations with the following time and spatial components:
\bea
\label{eom00}
{\alpha}^{3} \left(R_{0 0} \left[\hat{g}\right] - \frac{1}{2} {\hat{g}}_{0 0} R \left[\hat{g}\right] \right) &=& \frac{8 \pi G_0}{{c_0}^4} T_{0 0}~,\\
\label{eom11}
\alpha \left(R_{i i} \left[\hat{g}\right] - \frac{1}{2} {\hat{g}}_{i i} R \left[\hat{g}\right] \right) &=& \frac{8 \pi G_0}{{c_0}^4} T_{i i}~.
\eea

Some comments should be made here regarding the variational principle presented above. In our model the function $\alpha$ is a definite function of time. Such dependence may for example result from (\ref{govphi}) if the model is considered to be emergent from the theory of disformal gravity where we make the following identification: $\phi\equiv\alpha$. For the specific  ansatz for the time dependence of $\alpha(t)=\phi(t)$ the  Eq. (\ref{govphi}) can still be fulfilled at least locally (close to the singularity), provided that $f(\phi)$ and $g(\phi)$ were chosen properly and meet the condition implied by  (\ref{govphi}). Thus, the variation of the action (\ref{Action}) with respect  to the metric components $g_{\mu\nu}$ only should suffice to obtain the complete set of equations governing the evolution of the cosmological frame.

We assume the Friedmann metric for the matter field
\be
\label{FriedmanMetric}
ds_{M}^2= -{c_0}^2 dt^2 + a^2 (t) \left[ \frac{dr^2}{1-kr^2} + d{\theta}^2+ {\sin}^2 \theta d {\phi}^2 \right]~.
\ee
The resulting the gravitational metric ${\hat{g}}_{\mu \nu}$ takes the form:
\be
\label{NewMetric}
ds_{G}^2= -{c_0}^2 dt^2 + {\alpha}^2 a^2 (t) \left[ \frac{dr^2}{1-kr^2} + d{\theta}^2+ {\sin}^2 \theta d {\phi}^2 \right]~.
\ee
By inserting (\ref{FriedmanMetric}) and (\ref{NewMetric}) into the field equations (\ref{eom00}) and (\ref{eom11}) we obtain the density $ \rho \left( t \right) $ and the pressure $ p \left( t \right) $ in the following form
\bea
\label{Density}
\rho \left( t \right) &=& \frac{3 {\alpha}^3 \left( t \right) }{8 \pi G_0} \left( \frac{{ \dot{a}}^2 \left( t \right) }{a^2 \left( t \right) } + \frac{2 \dot{a} \left( t\right) \dot{\alpha} \left( t \right)}{ a \left( t \right) \alpha \left( t \right)} + \frac{ { \dot{\alpha}}^2 \left( t \right) }{{\alpha}^2 \left( t\right) } \right)~,\\
\label{Pressure} \nonumber
p \left( t \right)&=& - \frac{ {c_0}^2 \alpha \left( t \right)}{8 \pi G_0} \left( \frac{{ \dot{a}}^2 \left( t \right) }{a^2 \left( t \right) } + \frac{6 \dot{a} \left( t\right) \dot{\alpha} \left( t \right)}{ a \left( t \right) \alpha \left( t \right)} \right. \\
&+& \left. \frac{ { \dot{\alpha}}^2 \left( t \right) }{{\alpha}^2 \left( t\right) }   +  \frac{2 \ddot{a} \left( t \right)}{ a \left( t \right)} + \frac{2 \ddot{\alpha} \left( t \right)}{ \alpha \left( t \right)} \right)~.
\eea
The continuity equation is given by
\be
\label{ConEqu}
\dot{\rho} \left( t \right) + 3 \frac{ \dot{a} \left( t \right) }{ a \left( t \right) } \left( \rho \left( t \right) + \frac{{ \alpha}^2 \left( t \right)}{{c_0}^2}p \left( t \right) \right) + 3 \frac{\dot{\alpha} \left( t \right) }{\alpha \left( t \right)} \left( \frac{{\alpha}^2 \left( t \right)}{{c_0}^2} p \left( t \right) \right) = 0 ~.
\ee
Eqs. (\ref{Density}) and (\ref{Pressure})  are similar to the field equations in Brans-Dicke theory where  the gravitational constant is inversely proportional to the scalar field $\phi$ \cite{BransDickeTheory}. The function $\alpha$ plays the similar role as the scalar field $\phi$ in the Brans-Dicke theory. On the other hand, the conservation equation is different from the one derived in the Albrecht and Magueijo model \cite{AlbrehtMaguejo}, where an additional term contains time derivative of the dynamical gravitational constant $G(t)$ coupled to the density $\rho \left( t \right)$ - in our model the additional term contains the time derivative of $\alpha$ coupled to the pressure $ p \left( t \right) $. Let us notice that the assumption of  $ \alpha = const. $ does not lead to the Friedmann equations. This suggests that the gravitational constant and the speed of light are indeed dynamical parameters dependent on the instantaneous value of the $\alpha$ parameter, given by:
\bea
\label{gvarying}
G \left( t \right) &=& \frac{G_0}{{\alpha}^3 \left( t \right)}~, \\
\label{cvarying}
c \left( t \right) &=& \frac{c_0}{\alpha \left( t \right)}~.
\eea
This indicates that $ G_0 $ and $ c_0 $ appearing in Eqs. (\ref{Density}) and (\ref{Pressure}) have identical dimensions as their measured counterparts (\ref{gvarying}) and (\ref{cvarying}), but may have different values. Concluding, the assumption of $\alpha = const.$ and the interpretation of the gravitational constant and the speed of light as dynamical parameters given by (\ref{gvarying}) and (\ref{cvarying}), reduces the field equations (\ref{Density}), (\ref{Pressure}) and the conservation equation (\ref{ConEqu}) to standard set of cosmological equations with $G_{FM} = G_0 / {\alpha}^3$ and $c_{FM} = c_0 / \alpha$.

\section{Strength of singularities in dynamical fundamental constants theories}
\label{sec3}
\indent Hawking and Penrose were the first to properly define the singularity in general relativity. Their definition is based on the notion of geodesic incompleteness \cite{HavkingPenrose}. After discovering new types of the singular scenarios in cosmology \cite{FSF} characterised by different properties, Hawking-Penrose definition became no longer sufficient due to non-singular behavior of geodesics\cite{FernandezLazkos1,FernandezLazkos2}. In order to differentiate between the newly discovered types of the singular behaviours other criteria have to be introduced. Tipler and Kr\'{o}lak proposed criteria which are based on the notion of the ``strength'' of a singularity. In view of Tipler's definition \cite{TiplerDef}, the singularity is strong if the double integral diverges in a finite time:

\be
\label{tipler}
\int_0^{\tau} d\tau' \int_0^{\tau'} d\tau'' R_{\mu \nu}u^{\mu} u^{\nu} \rightarrow \infty ~,
\ee
where $ R_{\mu \nu} $ is the Ricci tensor, $ u^{\mu} $ is the 4-velocity, $\tau$ is the proper time. According to Kr\'{o}lak's definition \cite{KrolakDef}, the singularity is strong if the single integral diverges for the finite value of the parameter $\tau$:
%
\be
\label{krolak}
\int_0^{\tau} d\tau' R_{\mu \nu}u^{\mu} u^{\nu} \rightarrow \infty ~.
\ee
It is possible to compute the strength of the cosmological singularities generated by singular evolution of the dynamical fundamental constants in the bimetric framework introduced in Section \ref{sec2}. In such a model $ \alpha \left( t \right) $ is a part of the Ricci tensor and entails both the dynamical gravitational constant $G(t)$  and the varying speed of light $c(t)$ (Eqs. (\ref{gvarying}) and (\ref{cvarying})) .

We use Kr\'{o}lak or Tipler criterion to calculate the strength of the dynamical constants singularities. To compute (\ref{tipler}), (\ref{krolak}) we have to take into account the Ricci tensor calculated with the metric $ {\hat{g}}_{\mu \nu} $ and the 4-velocity $ {\hat{u}}^{\mu}=u^{\mu}= \left[ -1, 0, 0, 0 \right] $. The component $ R_{0 0} $ is given by
\be
\label{RicciTensor}
R_{0 0} =-\frac{ 3 \left[ 2 \dot{a} \left( t \right) \dot{\alpha} \left( t \right) + \ddot{a} \left( t \right) \alpha \left( t \right) + a \left( t \right) \ddot{\alpha} \left( t \right) \right]}{ a \left( t \right)\alpha \left( t \right)} ~.
\ee
For a non-singular scale factor in the range between $ 0 $ and $ t_s $ expressed by:
\be
\label{nonsingular}
a \left( t \right) = a_0 {\left( 1 + \frac{t}{t_s} \right)}^m ~,
\ee
where parameter $ m > 0 $ , we choose the function $ \alpha \left ( t \right) $ in the form
\be
\label{alpha1}
\alpha \left( t \right) = {\left(  1 - \frac{t}{t_s} \right)}^n ~,
\ee
where $n$ is some real number different than zero. For $ n > 0 $, the function $ \alpha \to 0 $ at $ t_s $. On the other hand,  for $ n < 0 $, the function $ \alpha \to \infty $ at $ t_s $. By inspecting the expression (\ref{RicciTensor}), we conclude that the assumptions (\ref{nonsingular}), (\ref{alpha1}) lead to a strong singularity in both cases, namely for  $ \alpha \to 0 $ and $ \alpha \to \infty $. Consequently,  we infer from (\ref{gvarying}) that the dynamical $ G \left( t \right) $  singularity is strong for $ G \left( t \right) \to 0 $ as well as for $ G \left( t \right) \to \infty $. The same refers to the dynamical speed of light (\ref{cvarying}). The singularity is likewise strong for $ c \left( t \right) \to 0 $ and for $ c \left( t \right) \to \infty $. An exception is the case n=1, where the singularity is strong with respect to Kr\'{o}lak's definition and weak with respect to Tipler's definition. By inserting (\ref{nonsingular}) and (\ref{alpha1}) into (\ref{Density}) and (\ref{Pressure}), we obtain for $ t \to t_s $ the following types of singular regimes:
\begin{itemize}
\item $\rho \left( t_s \right) \to \infty$ and $p \left( t_s \right) \to \infty$  for $n < 2/3$,
\item $\rho \left( t_s \right)\to \rho_s$ and $p \left( t_s \right) \to \infty$ for $n=2/3$,
\item $\rho \left( t_s \right) \to 0$ and $p \left( t_s \right) \to \infty$ for $2/3<n<2$,
\item $\rho \left( t_s \right) \to 0$ and $p \left( t_s \right) \to p_s$ for $n=2$,
\item $\rho \left( t_s \right) \to 0$ and $p \left( t_s \right) \to 0$ for $n>2$,
\end{itemize}
where $\rho_s$ and $p_s$ are some finite constants. It should be stressed that the above singular behaviour of the matter fields appear despite the perfectly regular behaviour of the metric associated with matter and are the consequence of the singularity that occurs in the causal structure defined by the gravitational metric.

A slowing down effect of the propagation of light was predicted by Quantum Loop Cosmology in the anti-new\-tonian limit, where the light stops to move as the energy density approaches the critical value \cite{quantumloopcosmology}.

It is also possible to calculate the ``strength'' of the singularity of the speed of light, but in this model the dynamical gravitational constant $ G \left( t \right) $ predominate over $ c \left( t \right) $, and the influence of the dynamical speed of light is reduced by the influence of the dynamical gravitational constant.
It should be stressed that the above results are generic and does not depend on the the particular dynamics of the Lorentz-violating field $\alpha (t)$ since they do not depend on the particular form of the bimetric model. Indeed, one does not have to specify the potential governing the dynamics of $\alpha (t)$ in order to perceive that the singular behaviour in the gravitational metric may lead to strong singularity in the matter sector even if the matter metric is perfectly regular. We find this particular feature of our study an advantage. The very fact that we do not give any particular equation of motion evolving the field $\alpha (t)$ does not weaken the main conclusions of our paper in any way.

\section{Conclusion}
\label{sec4}

We have shown that defining two the different causal structure leads to a dynamics which may in some aspects be similar to the dynamics of the scalar-tensor models. Such a result was obtained without specifying the kinetic sector of the theory that would determine the behaviour of $\alpha(t)$ which in our model is an assumed function of time. In the context of the scalar-tensor theories with disformal couplings such a behaviour of $\alpha(t)$ can, however, naturally emerge from the dynamical equation of motion for the scalar field $\phi$. The conformal factor which relates two different causal structures enters the field equation of the considered bimetric model in as similar way as the scalar field enters the field equations of the Brans-Dicke theory. The resulting field equations prove that the effective gravitational constant and the effective speed of light are the dynamical parameters and their values are related with the instantaneous value of the conformal factor.
It should be stressed that although the considered theory does not explicitly give the dynamics of the second metric $\hat{g}_{\mu \nu}$ (since there is no given the explicite form of the equation governing the dynamics of $\alpha(t)$) it sill constitutes a bimetric theory. Our model is based on the two different metrics: the one that describes the causal structure for the matter $g_{\mu \nu}$ and the other one that describes the causal structure for the gravitational field $\hat{g}_{\mu \nu}$. The fact that both metrics describe different causal structures is a direct consequence of the mathematical form of the action with its  gravitational part being computed on the metric $\hat{g}_{\mu \nu}$ (Eq. \ref{gravaction}) while its matter part is computed on the metric $g_{\mu \nu}$ (Eq. \ref{fieldaction}).
In the context of such a bimetric framework, we have found a particular scenario in which a strong singularity in the causal structure for the gravitational interaction in the sense of the criteria given by Tipler and Kr\'olak is accompanied by a singular behaviour in the matter fields despite the perfectly regular behaviour of the matter metric. In such a way, we have  proven that the singular behaviour of the matter fields does not necessarily have to be a result of an occurrence of the singularity in the causal structure for the matter, as it is the case for the most of the singular cosmological scenarios. \\

\textbf{Acknowledgments} We wish to thank Mariusz Dabrowski for discussions. This project was financed by the Polish National Science Center Grant  \\ DEC-2012/06/A/ST2/00395.

\end{document}